# Pursuing complex optical energy through Exceptional Points of Degeneracy (EPDs)

**SHAHRAM MORADI**[1, 2]

1 Electrical and Computer Engineering, University of Victoria, Victoria, Canada
2 Micro and Nano Technology Department, Middle East Technical University, Ankara, Turkey
Email: photonicsandoptics@gmail.com

We exploit balancing the complex optical energy between scattering and guiding states at contrived exceptional points of degeneracy in order to form an active waveguide without utilizing an active medium. This study reports a peculiar engineering of first order dispersion on a pseudo-Hermitian Hamiltonian system in which distributed Bragg grating tracks $0^{th}$ order of scattering phase-shift. This is owing to the EDPs instigated by means of 1D perturbing $\mathcal{PT}$ −symmetries in a planar photonic crystal. The coalescence of Bloch eigenstates occurs due to reverse amalgamation of modulated two-terminal optical component whose gain and loss parameters depends on the direction of light path. Then, we employed 1D binary superlittice around a defect that undergoes both states, physical equilibrium (in bound layer) and non-equilibrium (in unbound layer), in a chronological order. Then, recoupling optical energy to the underlying bound region compromises scattering phase-shift in each sequence. This achievement is not only broadening the understanding of revolutionized quantum mechanics but also benefits technological principles of steering localization with high Q-factors that are in huge demands of designing low-threshold switching, optical tweezers, lasing and various types of optical elements in photonic integrated circuits.

Introduction

Hermiticity and Poincar*é* group in physic describe discrete symmetry in quantum mechanics and the continuity respectively that produce a real spectrum for a Hamiltonian system. However, a $\mathcal{PT}$ −symmetry ensures the complex eigenvalues (if any) with a complex conjugate eigenvalue for its pair but does not guarantee to be Hamiltonian system[1]. According to the revolutionized quantum theories through some numerical[2,3] and experimental[1,4–11] results, being a Hamiltonian confirms a real spectrum of eigenvalues for even non-Hermitian with $\mathcal{PT}$ −symmetries. The validity of this concept spans from theoretical stage[12] to practical applications encompasses lasers[5], micro-nano resonators[4,7], NMR[13],

(meta)surfaces[9], nonlinear optics[10] and optical waveguides[4,6,7]. In photonics, some peculiarities via dexterous tailoring optical gain and/or loss parameters exploit non-Hermitian Hamiltonian systems even without resorting to $\mathcal{PT}$−symmetry[14] configurations. For instance, Kramers-Kronig optical media[15], imaginary gauge fields[16] and supersymmetric non-Hermitian photonics[17,18]. A difference between quasi-Hermitian and pseudo-Hermitian[19,20] bifurcates the $\mathcal{PT}$−symmetric components to two majors. In quasi-Hermitian, two identical irreducible sets (pairs) derived from physical characteristics of a system employs latter to fix metric operators[19]. However, in pseudo-Hermitian, contributing a third linear independent operator is required to compute inner products of two independent irreducible pairs with some common eigenvectors[19]. In perturbing problems[21,22], computing real eigenvalue spectra for broken symmetry[23,24] with a 1D perturbation effect[25] produces two irreducible Brillouin zones with independent eigenvectors. Since complete wave function depends on time, computing the scattering phase-shift[26–28] requires a third linear operator rather than two previous independent scattering matrices to pursue the envelope of optical energy in the Hilbert space. Here we suggest a lattice that undergoes a linear Berry phase via perturbing density of dielectric distribution per volume in which low-density possesses independent matrix basis of eigenvectors ($\widehat{\mathcal{M}}_N$) from the high-density volume ($\widehat{\mathcal{M}}_P$). Thus, for a special wave function such as $Ne^{ikr}$ in a chirality distribution, the momentum operator $\hat{p} = \frac{\hbar}{i}\frac{\partial}{\partial r}$ can give us a third set ($p = \hbar k$) as an interface between two independent sets of eigenvectors. A tangible pattern of a pseudo-Hermitian Hamiltonian system is suggested in which the inner products has a constant coefficient that can be obtained through iterative scheme ($\eta_+ \in \bigcap_{i=0}^{i=n}(\widehat{\mathcal{M}}_{N_i} \cap \hat{p}_i \cap \widehat{\mathcal{M}}_{P_i})$) between the metric operators including matrices of eigenvectors in each basis and momentum operators in their interface which plays a major role in the establishment of a complex dispersion contour. This is owing to Berry phase transition in a periodic lattice that provides a segment in broken $\mathcal{PT}$−symmetry that neither of eigenvectors in two pairs associated with it. In other words, the created exceptional points (EPs) produce coalescence instead of degeneracy due to an occurrence of only one eigenvector (e.g. either $\lambda = \lambda_1$ or $\lambda = \lambda_2$)[29]. Therefore, eigenvectors in the exceptional points (EPs) get a superposition of these two pairs that undergoes an abrupt phase transition in each adiabatic process. The linear operator with a $\eta$−pseudo-Hermitan satisfies being a Hamiltonian system shown with the identity operator ($\eta$), if $H^\dagger = \eta H \eta^{-1}$. In other words, breaking $\mathcal{PT}$-symmetry in each cycle of geometrical transition must result in constant coefficient to produce a real spectrum. Thus, the computed values from two dimensions (reduced from high dimensions)[29,30] guarantee the real spectrum only in close vicinity of Exceptional Points of Degeneracy due to vanishing norm (scalar product with self-orthogonality)[29,31,32].

**Balancing the chromatic dispersion**

By tuning a physical parameter (here perturbing the density of the dielectric distribution) engineering the dispersion (time-variant)[33,34] is exploited by means of periodic asymmetric pairs of clusters. The suggested asymmetric lattice undergoes an abrupt phase transition that produces an excess shift in phase (either delay or advance) due to the proportionality of probability distribution to the energy difference in heterojunctions. We compute real eigenvalues and consequently eigenmodes for such pseudo-Hermitian Hamiltonian configuration. The suggested periodic lattice systematically divides the perturbation into two types: low to high (LH) density and high to low (HL) density of distributing dielectric through light flow. The effect of the geometric perturbation on abrupt phase transition evolves a wave packet to change superposition and possess a group velocity either slower than light (STL) or faster than light (FTL). To elaborate it further, we apply perturbation effect on a periodic all-dielectric photonic crystal to modify the density of states (DOS) for the addressed p-polarized (TM) radiated wave in $\Gamma M$ direction. The applied perturbing breaks the symmetric potential wells in the planar photonic crystal and alter Bloch envelop based on light flow direction. According to the study in reference [25], the perturbing density of dielectric distribution in grating structure reproduces an inconstant slope of wave-vector $k'(\omega)$ corresponding to the two generated eigenstates in the first and second degrees of degeneracies. In chromatic dispersion, the gradient of wave-vector, $k'(\omega, \vec{r})$, determines wave-packet either to undergo loss ($k'(\omega, \vec{r}) > 0$) or gain ($k'(\omega, \vec{r}) < 0$) in terms of optical energy. To prove such transition in phase, we employ frequent reverse amalgamations of the module in a 1D binary superlattice. Then we discover a broadband $0^{th}$ order in k points [25,35–43] that is in accordance with computed dispersion diagram (see the results in supplementary information). In conventional photonic band gap, however, such stop band states are vulnerable due to the nature of phase dispersion that vanishes coupling modes confining in an embedded defect (light path), but the $0^{th}$ order of photonic band gap based on alternating stacks with optical gain and loss produce a complex optical energy with a scattering phase-shift[26–28].

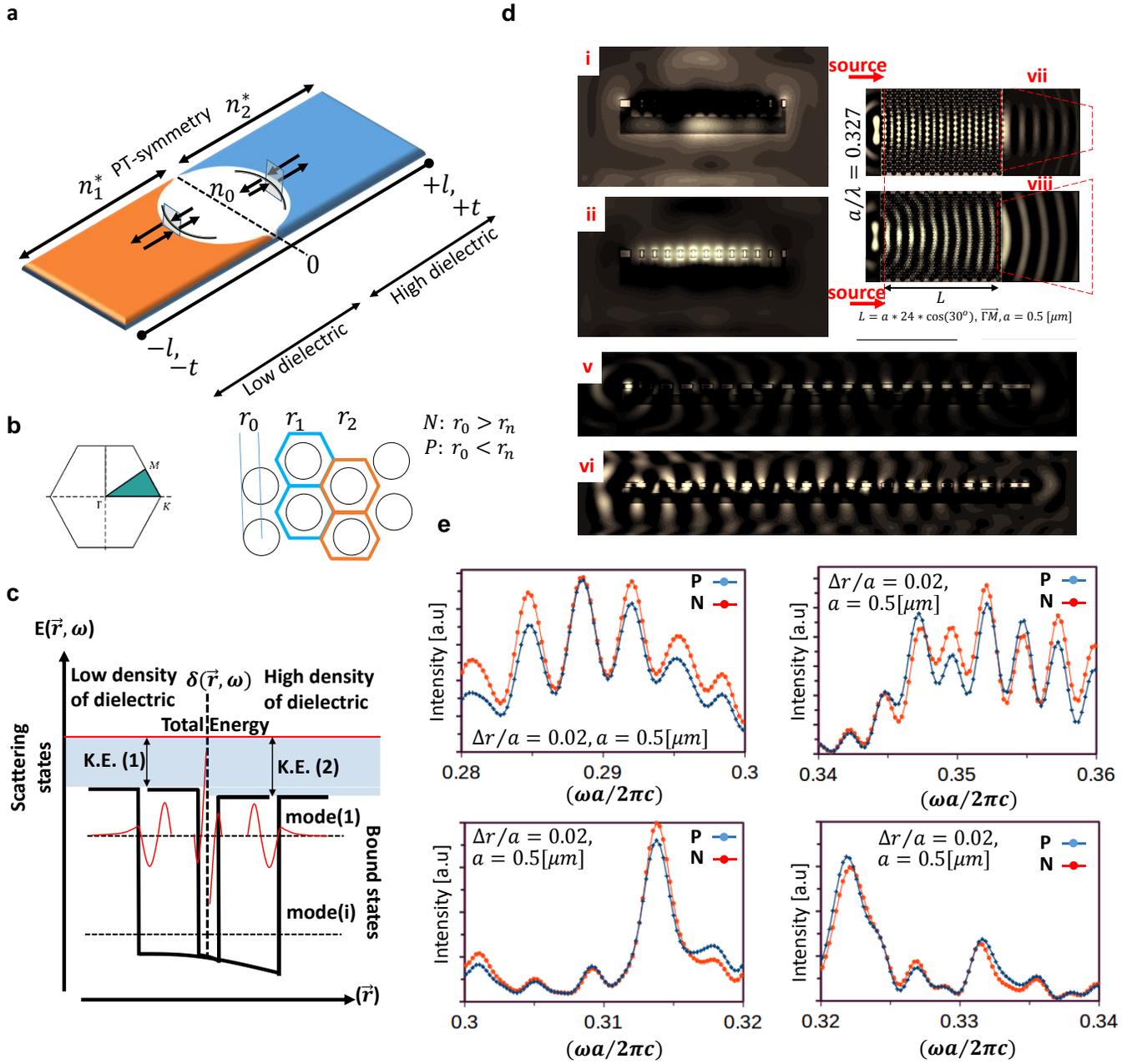

Fig. 1 | **Introducing the fundamentals of the two-terminal optical component. a.** Schematic of $\mathcal{PT}$−symmetry formation due to Berry phase transition in each side (orange and blue) based on the perturbing lattice via density of dielectric distribution which causes asymmetric boundary interfaces with respect to the normalized vacuum index ($n_0$) and two complex refractive index ($n^* = n + j\kappa$ or $|n|\angle\varphi$) are adjoint of each other. **b.** A hexagonal lattice and one type of perturbing parameter (radii) that forms P and N components based on Berry phase transition. **c.** A band diagram for asymmetric distribution of dielectric that causes perturbation on the depth of potential wells and formation of Delta function $\delta(\vec{r}, \omega)$ which depends on the direction in the reciprocal lattice and on the frequency of localized wave. **d.** Field intensity map (using linear grayscale) in the different structures. Insets: The scattering states in a right hand side material that causes de-localizing (loss) shown in (**i, v**). The

bound states in a left hand side material shown in **(ii, vi)** that causes localizing (gain) optical energy. The **(vii, viii)** shows how chromatic dispersion in a complex refractive index $n^* = n + j\kappa$ or $|n|\angle\varphi$ in which the quantum coherency (up) and de-coherency (down) for an applied pure state of light pulse (Gaussian pulse) follows establishing state in the medium. In the **(vii)** bound state stablishes left hand side material (LHM) in which $\varphi < 0$ due to $k'(\omega, \vec{r}) < 0$. In the **(viii)** scattering state establishes the right hand side material (RHM) in which $\varphi > 0$ due to $k'(\omega, \vec{r}) > 0$. **e.** The transmission spectrum for two evolved Gaussian light pulses in one component with a mild perturbation ($\Delta r = 20[nm]$, $a = 0.5[\mu m]$) from two reverse directions that illustrates indeterminate difference in amplitude and/or gradient of magnitude capable of forming different interference effect in superposition and/or probability distribution in the vicinity of exceptional points (EPs).

Thus, we exploit this optical component to form an active waveguide without utilizing the gain medium. Unlike the traditional passive waveguides with a temporal localized optical energy, decaying based on the dependency of a reflection phase of each mirroring layer (fiber Bragg gratings) on angle[44], the modulated mirroring layers with zero phase delay[36] eliminates the dependency of the reflection phase on reflection angle. So, under continuous-lengthy waveguide-condition, applied Gaussian pulse (e.g. from a fiber coupled diode laser) evolves from a small effective mode area (both active and passive) into an intense gaining optical energy with standing waves.

In Fig. 1-a, a schematic of $\mathcal{PT}$-symmetry component with a pseudo-Hermitian Hamiltonian configuration is designed in which perturbing effect causes the arguments of complex refractive index to become either $\varphi < 0$ or $\varphi > 0$ and consequently the establishment of LHM or RHM respectively. This is owing to varying the density of dielectric distribution (Fig. 1-b) in 1D orthogonal basis of hexagonal lattice and gives rise to opposite chromatic dispersions ($k'(\vec{r}, \omega) < 0, k'(\vec{r}, \omega) > 0$). Thus, the band diagram with an asymmetric potential well between two irreducible Brillouin zones grows with a narrow edge (delta function shown in Fig. 1-c). To satisfy the continuity of the wave function in each interface and normalizing it in infinity, the solution in the narrow edge has to recoil abruptly from scattering states to bound states or vice versa. The momentum variation in kink shape of two wave functions follows energy differences with elapsing time in each perturbing section. Thus, accumulation of scattering phase-shift after frequent rebounding (switching between localizing gain/loss) is expected (Fig. 1-d) for the superposition of two states. A superposition with the probability density for two oscillating wave functions $|\Psi_a(\bm{r},t) + \Psi_b(\bm{r},t)|^2$ at the angular frequency of $\omega_{ab} = |E_a - E_b|/\hbar$ in which the energy differences does matter for the evolution and it is shown in Fig. 1-e. In Fig. 2, superposition of two states with different orders (LH or HL) gives rise to a comparable scattering phase-shift for particular angular frequencies that is shown with blue transparent bands.

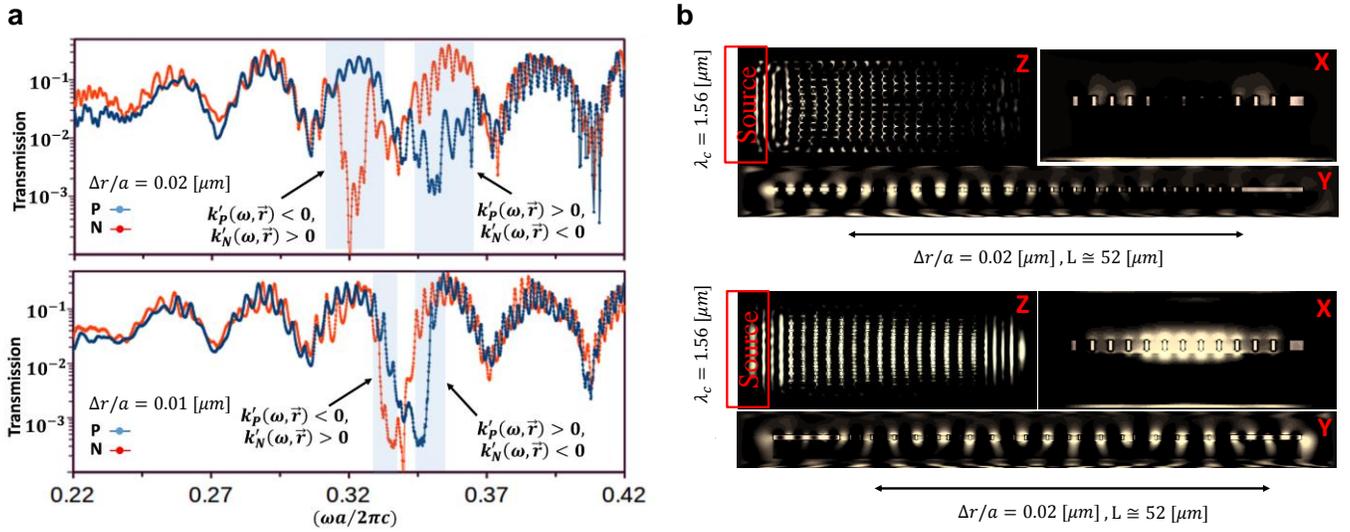

Fig. 2 | **Normalized transmission spectrum in logarithm scale and the field intensity map (using linear grayscale)**. **a.** Two components that exhibit symmetric phase transition (scattering phase-shift) in a particular range of frequencies (transparent blue bands) due to being opposite chromatic dispersion through P component (blue) and N component (red). **b.** Applying light pulse with $f_c = 0.32$ and width of $\Delta\omega = 1E-1$ (angular frequency profile) to two components (P component in up and N component in down) with their side views in "z", "x" and "y" that illustrate two suggested components with a pseudo-Hermitian Hamiltonian configuration that have a comparable probability distribution for an evolved superposition in certain bands.

## Active resonated waveguide

The light-base technology utilizes the resonators for optical amplification not only stand-alone but also in any integration setup such as mode locking, wavelength tuning, switching components or optical modulators. Exploiting nonlinearity effect of crystal materials based on gaining medium such as lithium niobate [45] exploitation in photonic integrated circuits (PICs) is an ongoing option for an active waveguide. Confining light in the waveguide with no contribution of gaining medium that we discuss in this study with elaborate manipulating of first and higher orders of dispersions in a light path surrounded by the photonic crystal. Thus far, we designed a 1D binary superlattice (see the supplementary information) with a $0^{th}$ order of phase reflecting that is capable of constructive evolving the wave-packet.

One of the considerations in designing waveguide has to do with a high-quality factor for resonated modes. However, the governing principle to reach such goal relies on reflection angle in mirroring layers that determines the superposition of ultimate envelope. One of the characteristics of zero index material is a $0^{th}$ order of phase reflection that accentuate its peculiarity in terms of application. In other words, chromatic dispersion regulates the angle of the radiation and consequently interference of the coupling modes. Thus, supercoupling effect[46] as a unique artificial optical property provides an exotic platform to produce a constructive resonance.

For the sake of comparison, we applied a broad Gaussian pulse to excite all resonated modes in Slab, P component, N component and superlattice of P-N pairs that are shown in (Fig. 3).

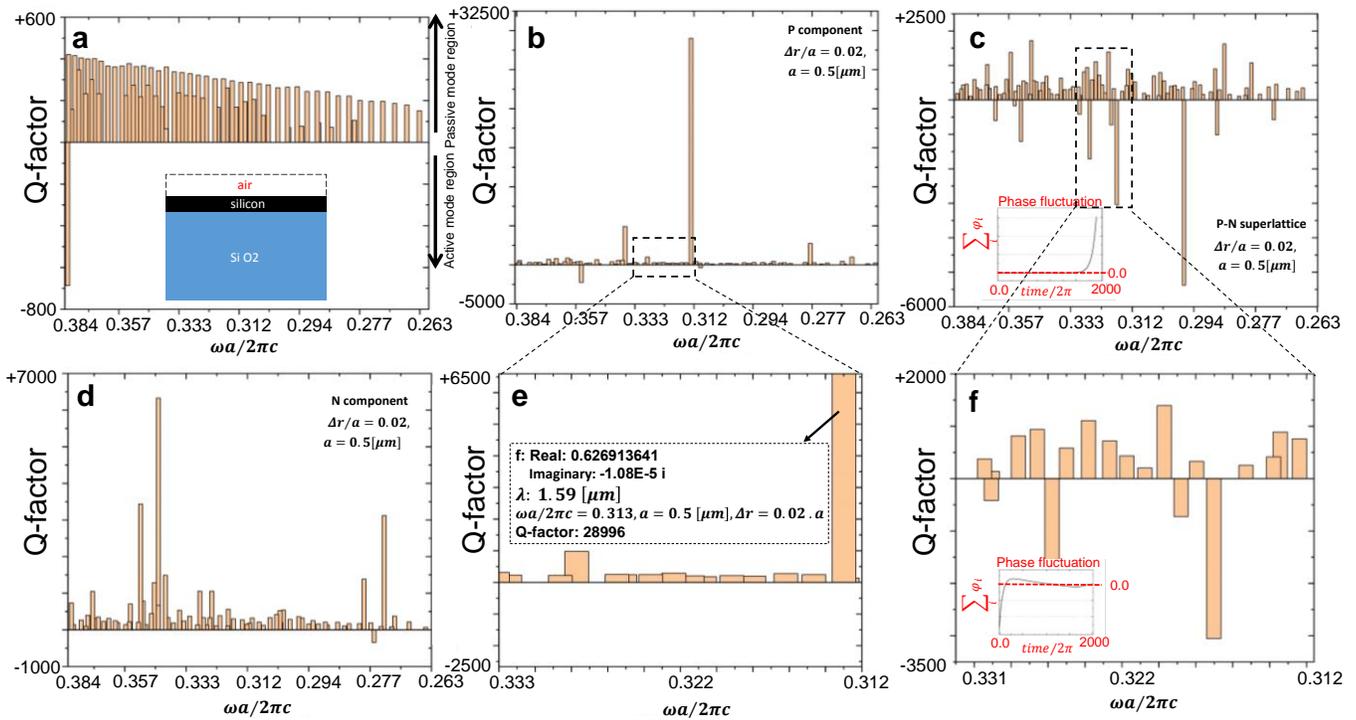

Fig. 3 | **Resonated modes with complex frequencies (if any) that are shown with real parts of angular frequencies versus Q-factors. a.** In a pure slab (silicon) on top of (silicon dioxide) as a substrate, **b.** In **P** component with Berry phase transition ($\Delta r = 10\ [nm]$) in whole geometrical cycles of the lattice, **c.** In **P-N** superlattice with balancing gain and loss parameters for obtaining zero-$\hat{n}$ gap within $\Delta\omega. a/2\pi c = 0.305{\sim}0.335$ range; inset: Indicate fluctuation of all modes does not satisfy the $0^{th}$ order of phase reflection, **d.** in **N** component with Berry phase transition ($\Delta r = 10\ [nm]$) in whole geometrical cycles of the lattice, **e.** A passive mode with a high impact factor within the $\Delta\omega. a/2\pi c = 0.305{\sim}0.335$ range, **f.** Compensating passive (loss) and active (gain) resonated modes to create a $0^{th}$ order of phase fluctuation within the $\Delta\omega. a/2\pi c = 0.305{\sim}0.335$ range that is shown in inset (see the supplementary information for creation of a zero-$\hat{n}$ gap in this range of angular frequencies) that satisfies the $0^{th}$ order of phase reflection.

Since the first order dispersion is a function of frequency, we tune the design parameter for the $0^{th}$ order in a way to achieve gaining resonated modes in a particular range of wavelength in $1.45\ [\mu m]{\sim}1.65\ [\mu m]$, which corresponds to an angular frequency around $\Delta\omega.a/2\pi c = 0.305{\sim}0.335$. According to the different suggested structures, the formation of active resonated modes is unlikely possible unless the tailoring design possesses a balancing between gain and loss parameters. Although the Q-factor shown in Fig. 3-b reaches to 28996 (with more details in Fig. 3-e) but the creation of active mode does not occur in this region (see the selected details about the number of modes, maximum Q-factor, number of active/passive modes and crude errors of computation in Table 1). In Table 1, the explanation of excited modes for each module illustrates complex frequencies (if any) that propagate at the termination of each composite. As it is clear from Table 1, the number of resonating modes increases in the $0^{th}$ order of mirroring layer due to less destructive interaction of modes. Furthermore, the number of active modes with positive imaginary part of resonating modes are positive is more than the other designs (pure slab, P and N components). To interpret computed modes (see the MEEP documentation, discussion in chapter "resonant modes and transmission in a waveguide cavity"), we need to distinguish guided modes from those of leaky or extended modes. Computing band structure for each configuration will clarify the region corresponded to extended modes that are not below the light cone ($\omega > ck_r$) from those guided modes that are below the light line. The discrete bound states in the dispersion diagram provide a reliable map to figure out the guided bands that are in bound state (localized) to the impinging light path through the waveguide. Thus, the imaginary parts for computed angular frequencies with a negligible value in the first and second bands is a guided mode. According to the dispersion diagram, the existence of bands rather than first and second one indicates some leaky modes due to the numerical errors such as the finite computational cell size that causes overlapping guided modes to the perfect match layer (PML). These leaky modes are artifacts based on PML boundaries that do not absorb light while propagating in an excellent way. In addition, reporting loss per unit distance for a suggested waveguide is a critical matter that can be computed via dividing loss (twice the imaginary part of given complex angular frequency) per unit time. The slope of bound states with respect to the wavevector ($k_r, \vec{r} \to \Gamma M$) determines time scale of a given mode that is acquired through ($d\omega/dk_r$), therefore, examining localized modes through the waveguide can be achieved through the lifetime of each computed mode that is given with Q as a threshold.

Table 1 | Characteristics of some selected mode in different design and structures

| Type of media | Number of possible excited modes | Modes (Gain/Loss) with the highest optical periods ($\omega/2\pi c$, $a = 0.5\ [\mu m]$) | | Modes with Gain ($\omega/2\pi c$, $a = 0.5\ [\mu m]$) | | | |
|---|---|---|---|---|---|---|---|
| | | Complex angular frequency | Q factor | Complex angular frequency | Corresponding wavelength [$\mu m$] | Q factor | Crude numerical error |
| Slab | 71 | 0.760 + 5.54E-04 j | -685 | 0.760 + 5.54E-04 j | 1.310 | -685 | 1.65E-05 |
| | | 0.765 - 8.90E-04 j | +429 | | | | |
| Waveguide surrounded with **P** components | 75 | 0.553 + 8.01E-04 j | -345 | 0.553 + 8.01E-04 j | 1.800 | -345 | 1.83E-05 |
| | | 0.690 – 2.60E-04 j | +1289 | 0.525 + 9.94E-04 j | 1.900 | -264 | 1.37E-04 |
| Waveguide surrounded with **N** components | 75 | 0.709 + 1.50E-04 j | -2246 | 0.709 + 1.5E-04 j | 1.410 | -2246 | 5.23E-04 |
| | | 0.626 – 1.08E-05 j | +28996 | 0.620 + 8.3E-04 j | 1.610 | -372 | 5.47E-07 |
| Waveguide surrounded with supperlattice of **P-N** components | 94 | 0.696 – 2.02E-04 j | +1721 | 0.709 + 2.1E-03 j | 1.410 | -167 | 6.56E-05 |
| | | | | 0.704 + 2.9E-04 j | 1.420 | -1202 | 1.12E-05 |
| | | | | 0.660 + 7.9E-04 j | 1.515 | -416 | 7.13E-06 |
| | | | | 0.653 + 1.9E-04 j | 1.530 | -1708 | 6.11E-04 |
| | | 0.593 + 5.50E-05 j | -5377 | 0.638 + 4.4E-04 j | 1.565 | -724 | 6.15E-04 |
| | | | | 0.634 + 1.0E-04 j | 1.575 | -3052 | 4.94E-03 |
| | | | | 0.593 + 5.5E-05 j | 1.690 | -5377 | 6.91E-04 |
| | | | | 0.575 + 2.8E-04 j | 1.740 | -1014 | 6.79E-05 |

The phase fluctuation rate in the subset of Fig. 3-**f**, illustrate overall and average phase accumulation ($\sum_i \varphi_i$) for P-N superlattice that becomes a net near to the zero for particular range of frequencies. This is owing to the designing parameters (alternating gain and loss) with a reverse combination of modulated two-terminal optical component.

In Fig. 4, the full-wave analysis presents near field solutions for the different suggested components. Fig. 4-**a** shows an optical frequency comb as an input source consists of a series of discrete, partially equal spaced frequency, lines that characterize the Gaussian pulse in the Fourier domain. The wave packet is localized in the output of silicon slab (shown in Fig. 4-**a**) oscillating (breathe) in time along with decaying amplitude. A stabilized pulse train with a constant phase that can be seen through the constant distances of the parallel tracks of lines in polarization projection in Fig. 4-**a**. The partially polarized light pulse Fig. 4-**d** undergoes a polarization rotation (clockwise Fig. 4-**e**) during the light flow. Introducing this optical pulse to the suggested planner photonic crystal with an embedded slice of defect in the middle that is shown in Fig. 4-**f** provides a promising results for our suggested structures. Here we only demonstrate a finite number of iteration up until reaching the amplitude to the third decades below the maximum magnitude in first generated pulse. First, we apply this optical pulse to the waveguide surrounded by N components in which a mode (0.620 + 8.3E-04j) possesses a

Q factor around +28996, high enough to form an envelope with a high magnitude in resonated region. The result for near field output in Fig. 4-b reveal a primary overshooting in time, but decaying after reaching to the $\cong 1200$ optical periods. Meaning that despite the resonating mode through the waveguide the wave packet distortion through the passive optical elements are inevitable. Finally, applying 1D binary superlattice provides an active resonance in the output of a waveguide in which the gain and loss parameters for chromatic dispersion are in balanced. The continuous and ascendant in magnitude of wave packet in Fig. 4-c demonstrate a coherence transition of light pulse due to the $0^{th}$ order of dispersion. A balanced chromatic dispersion including P and N components in which an engineering dispersion provides a promising platform to exploit a unique optical phenomenon, the $0^{th}$ order of dispersion, in a broadband region. In a word, the complex nature of electromagnetic with a dexterous engineering of artificial matter brings such peculiarity, which are in a high demands in all advanced light-based technologies.

Here to form a huge contrast in the density of air volume in a particular direction, we apply two approaches simultaneously. First, the idea of compressing cylindrical holes in the $\Gamma M$ direction attuned to $35^o$ (see the idea in[25]). Second, the diameter and height of the cylindrical air varies in a way to provide huge contrast of dielectric volume in two layers. As a result, perturbation of geometry arranging from low to high (LH) density of volume provides a particular frequency band spectrum that all have a common mismatch in phase.

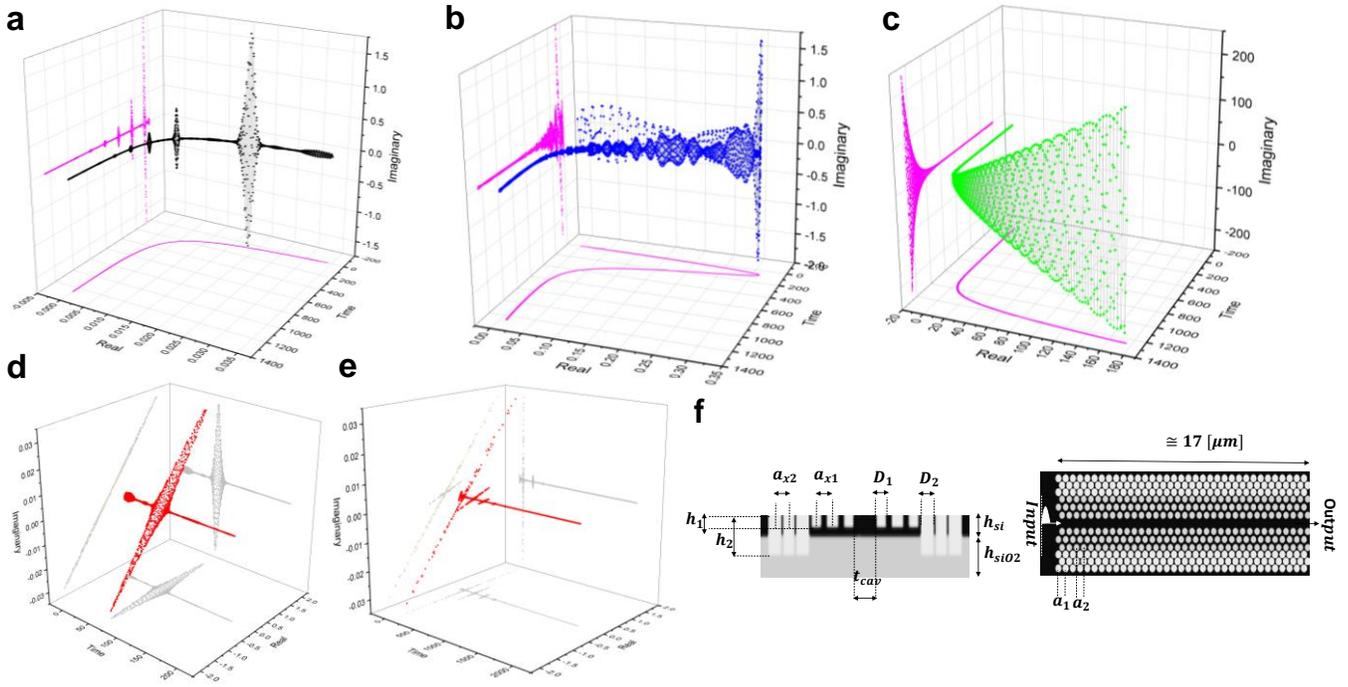

Fig. 4 | **Electromagnetic full wave analysis in P-N superlattice. a.** The wave packet transition in a slab waveguide in which polarization remains constant but the intensity decays in magnitude in harmonic periods. **b.** The wave packet distortion in a passive waveguide with high to low (HL) density of distributing dielectric representing P component in which temporary overshooting intensity profile occurs due to a passive high Q-factor resonance. **c.** The wave packet transition in an active waveguide with the $0^{th}$ order of phase dispersion that is established in a pseudo-Hermitian Hamiltonian configuration in which the gain and loss parameters are balanced for a particular range of frequencies. **d.** The partially polarized light pulse as a reference in slab for normalizing the transmission spectrum. **e.** The extended polarization profile for normalizing the wave packets in lengthy structures. **f.** A rough example of creating a contrast in the density of dielectric per surface volume with a lattice constant equal to $a_1 = a_2 = a = 0.5\ [\mu m]$; $hsi = 0.32\ [\mu m]$, $hsiO2 = 1\ [\mu m]$, $h1 \cong 0.6\ hsi$, $h2 \cong 0.6\ hsio2$, $D1 = 0.35\ [\mu m]$, $D2 = 0.4\ [\mu m]$, $a_{x1} = a.\cos(\pi/6)\ [\mu m]$ as a regular hexagonal lattice, $a_{x2} = a.\cos(\pi * 0.195)\ [\mu m]$ as a compressing irregular perturbed lattice and the dimension of the slab waveguide is $t_{cav} = 2 * a_{x1}$ with the $\cong 17[\mu m]$ length.

---

Due to the opposite characteristics of dispersion in a range of frequency, we expect the creation of zero-n gap with zero phase delay. Thus, the resonated modes with zero phase transition provides a unique platform to obtain a strong light matter interaction due to freezing wave packet with zeroth order of group velocity.

## Conclusion

This study introduces for the first time a two-terminal optical component with odd complex effective refractive index in two reverse 1D directions. We modulate the two-terminal optical component via chirality in distributing density of dielectric with applying a Berry phase transition on a planar photonic crystal. The kinetic energy (scattering state) of impinging electromagnetic waves undergoes scattering phase-shift based on breaking parities via chirality in the density of distributing the dielectric per surface volume. Therefore, an exotic dispersion diagram encompasses odd group velocities ($|k'_1(\omega,\vec{r})| = |k'_2(\omega,\vec{r})|$) forms directionality in the chromatic dispersion. Then, we acquired a balanced optical system with a pseudo-Hermitian Hamiltonian configuration that possesses the $0^{th}$ order of phase dispersion. In conclusion, we exploited it to form an active waveguide without utilizing an active medium. Unlike the traditional passive waveguides that resonates temporarily, the proposed balanced layer enhances wave function due to $0^{th}$ order of phase dispersion. By considering today's communication substructure that faces a serious scalability and energy consumption issues, reported results in this study would be a novel and unique approach capable of dramatic modification to both fundamental working principles in quantum mechanics and saving power budgets for optical data transmitting in light-based devices.

## Methods

**Programming tools:** We used MIT Photonic-Bands (MPB[47]) as a well-known eigenmode solver via Plane Wave Expansion (PWE) method to reach the accurate photonic band structure that provides a connection between given electromagnetic radiation and optical medium properties. In addition, using FDTD method via MIT electromagnetic equation propagation (Meep[48]), provides all results for field intensity maps with a linear grayscale, transmission spectrums and field components along with decay rate of modes. We used the time series $f(t) = \sum_n a_n . e^{-i\omega_n t}$ corresponding to the extracted E-field components at the end of the resonator with applying a broadband Gaussian pulse to flow through the suggested waveguide. We analyzed all excited modes to compare active and passive modes in the given structure. Here we performed a data analysis to form a 3D full-wave propagation in the time domain for a visualized examination of the phase, the amplitude, and the polarization of modes. Since we extract 6 columns of data in which two columns represent complex values of resonated frequencies, two other columns give complex values of decaying rate and one column of data provides $Q$ number that is the number of optical periods of resonated modes until the decaying by the exponential terms.($Q = \frac{Re\ \omega}{-2\ Im\ \omega}$)

In the Fourier domain, we achieved fractional bandwidth at half-maximum of the resonated modes with the reverse quantity of quality factor ($Q^{-1}$). In addition, the uncertainty in the processing of all extracted data is showing by an error at the end of each characterized mode. These errors are not referring to the miscalculation of size and/or solution of given geometrical parameters. It worth to mention that the '*harmiv*' command in Meep[48] extracts all the data after the source is switched to the off state at the end of the waveguide.

**Solutions:** By considering the linearity of Maxwell's equations in the perturbed interface for both fields in time, $E(r, t) = E(r)e^{j\omega t}$ and $H(r, t) = H(r)e^{j\omega t}$, the solution via Fourier theory will not construct the result for its given components. Solving master-equation for both fields give rise to a combination of propagation modes as a solution for given $\omega$ through the band structure [49]. Here we use master equation (eigenvalue equation[50]) instead of time-independent Schrödinger equation[51] to compute eigenvalues for all possible eigenstates: $\Theta H(r) = \left(\frac{\omega}{c}\right)^2 H(r)$

Where $\Theta H(r) \equiv \nabla \times \left(\frac{1}{\varepsilon(r)} \nabla \times H(r)\right)$, exhibits a functional form of dielectric to determine the magnetic field. Since the Hermitian ($\Theta$) operates just on the magnetic field due to the Hermiticity of the magnetic flux density. Thus, the determination of the electric field will be the next computing step after obtaining **H** field. Therefore, applying a translation of any symmetry such as a photonic crystal gives rise to the dispersion relation in which the splitting bands in each degree of degeneracy occurs.

**Numerical and analytical results:** Implementing Bloch's theorem via MPB[47] computational tools, we conduct a simulation with a high-resolution FE method in which the grid elements is applied for 3 dimensions structure in a hexagonal lattice distribution of cylindrical holes. We depicted the computed result of photonic band structure for such lattice in a perturbed form. The lattice characterizes with the perturbed periodic pairs to compute its transmission spectrum, the field intensity, and the effective refractive index. We calculate only TM-mode that is corresponded to the eigenmodes of the z-component for the applied electric field ($E_z$) through the proposed lattice. We summarized reduced basis of lattice in three directions (**Γ**, **M** and **K**) due to 2-dimensional symmetries in the first Brillouin zone.

### Data availability

We provided all files in (....google drive) along with the manuscript for reviewers along with the both manuscripts including excel data and scheme programming manuscripts.

## References

1. Konotop, V. V., Yang, J. & Zezyulin, D. A. Nonlinear waves in PT-symmetric systems. *Rev. Mod. Phys.* **88,** 035002 (2016).
2. Bender, C. M. & Boettcher, S. Real Spectra in Non-Hermitian Hamiltonians Having P T Symmetry. *Phys. Rev. Lett.* **80,** 5243–5246 (1998).
3. Bender, C. M., Boettcher, S. & Meisinger, P. N. PT-symmetric quantum mechanics. *J. Math. Phys.* **40,** 2201–2229 (1999).
4. Regensburger, A. *et al.* Parity–time synthetic photonic lattices. *Nature* **488,** 167–171 (2012).

5. Feng, L., Wong, Z. J., Ma, R.-M., Wang, Y. & Zhang, X. Single-mode laser by parity-time symmetry breaking. *Science* **346,** 972–5 (2014).

6. Makris, K. G., El-Ganainy, R., Christodoulides, D. N. & Musslimani, Z. H. Beam Dynamics in P T Symmetric Optical Lattices. *Phys. Rev. Lett.* **100,** 103904 (2008).

7. Peng, B. *et al.* Parity–time-symmetric whispering-gallery microcavities. *Nat. Phys.* **10,** 394–398 (2014).

8. Painter, O. *et al.* Two-dimensional photonic band-Gap defect mode laser. *Science* **284,** 1819–21 (1999).

9. Fleury, R., Sounas, D. L. & Alù, A. Negative Refraction and Planar Focusing Based on Parity-Time Symmetric Metasurfaces. *Phys. Rev. Lett.* **113,** 023903 (2014).

10. Dmitriev, S. V., Sukhorukov, A. A. & Kivshar, Y. S. Binary parity-time-symmetric nonlinear lattices with balanced gain and loss. *Opt. Lett.* **35,** 2976 (2010).

11. Suchkov, S. V. *et al.* Nonlinear switching and solitons in PT-symmetric photonic systems. *Laser Photon. Rev.* **10,** 177–213 (2016).

12. O'neill, B. *Semi-Riemannian geometry with applications to relativity*. (1983).

13. Galda, A. & Vinokur, V. M. Parity-time symmetry breaking in magnetic systems. *Phys. Rev. B* **94,** 020408 (2016).

14. Letters), S. L.-E. (Europhysics & 2018, undefined. Parity-time symmetry meets photonics: A new twist in non-Hermitian optics. *iopscience.iop.org*

15. Ahmed, W. W. *et al.* Directionality Fields generated by a Local Hilbert Transform. (2017). doi:10.1103/PhysRevA.97.033824

16. Hatano, N. & Nelson, D. R. Localization Transitions in Non-Hermitian Quantum Mechanics. *Phys. Rev. Lett.* **77,** 570–573 (1996).

17. Longhi, S. Invisibility in non-Hermitian tight-binding lattices. *Phys. Rev. A* **82,** 032111 (2010).

18. Teimourpour, M. H., Ge, L., Christodoulides, D. N. & El-Ganainy, R. Non-Hermitian engineering of single mode two dimensional laser arrays. *Sci. Rep.* **6,** 33253 (2016).

19. Mostafazadeh, A. Pseudo-Hermitian Representation of Quantum Mechanics. (2008). doi:10.1142/s0219887810004816

20. Mostafazadeh, A. & Batal, A. Physical aspects of pseudo-Hermitian and *PT*-symmetric quantum mechanics. *J. Phys. A. Math. Gen.* **37,** 11645–11679 (2004).

21. Lee, T. E. & Joglekar, Y. N. PT -symmetric Rabi model: Perturbation theory. *Phys. Rev. A* **92,** 042103 (2015).

22. Znojil, M. Admissible perturbations and false instabilities in PT -symmetric quantum systems. *Phys. Rev. A* **97,** 032114 (2018).

23. Jin, L. Parity-time-symmetric coupled asymmetric dimers. *Phys. Rev. A* **97,** 012121 (2018).

24. Afzal, M. I. & Lee, Y. T. Supersymmetrically bounding of asymmetric states and quantum phase transitions by anti-crossing of symmetric states. (2016).

25. Moradi, S., Govdeli, A. & Kocaman, S. Zero average index design via perturbation of hexagonal photonic crystal lattice. *Opt. Mater. (Amst).* (2017). doi:10.1016/j.optmat.2017.09.008

26. Rașeev, G. & Atabek, O. Resonant behaviour of the scattering phase shift. *Nuovo Cim. B Ser. 11* **107,** 463–481 (1992).

27. Nowakowski, M. & Kelkar, N. G. The Use of the Scattering Phase Shift in Resonance Physics. (2004). doi:10.1142/9789812701855_0027

28. Prelovsek, S., Lang, C. B. & Mohler, D. Scattering phase shift and resonance properties on the lattice: an introduction. (2011).

29. Heiss, W. D. The physics of exceptional points. *J. Phys. A Math. Theor.* **45,** 444016 (2012).

30. Heiss, W. D. & Harney, H. L. The chirality of exceptional points. *Eur. Phys. J. D* **17,** 149–151 (2001).

31. Krejčiřík, D., Siegl, P., Tater, M. & Viola, J. Pseudospectra in non-Hermitian quantum mechanics. *J. Math. Phys.* **56,** 103513 (2015).

32. Narevicius, E., Serra, P. & Moiseyev, N. Critical phenomena associated with self-orthogonality in non-Hermitian quantum mechanics. *Europhys. Lett.* **62,** 789–794 (2003).

33. Caloz, C. Metamaterial Dispersion Engineering Concepts and Applications. *Proc. IEEE* **99,** 1711–1719 (2011).

34. Zurita-Sánchez, J., Halevi, P., A, J. C.-G.-P. R. & 2009, undefined. Reflection and transmission of a wave incident on a slab with a time-periodic dielectric function. *APS*

35. Chen, W.-J., Hou, B., Zhang, Z.-Q., Pendry, J. B. & Chan, C. T. Metamaterials with index ellipsoids at arbitrary k-points. *Nat. Commun.* **9,** 2086 (2018).

36. Kocaman, S. *et al.* Zero phase delay in negative-refractive-index photonic crystal superlattices. *Nat. Photonics* **5,** 499–505 (2011).

37. Mocella, V. *et al.* Self-Collimation of Light over Millimeter-Scale Distance in a Quasi-Zero-Average-Index Metamaterial. *Phys. Rev. Lett.* **102,** 133902 (2009).

38. Hao, J., Yan, W. & Qiu, M. Super-reflection and cloaking based on zero index metamaterial. *Appl. Phys. Lett.* **96,** 101109 (2010).

39. Liberal, I. & Engheta, N. The rise of near-zero-index technologies. *Science* **358,** 1540–1541 (2017).

40. Li, Y. *et al.* On-chip zero-index metamaterials. *Nat. Photonics* **9,** 738–742 (2015).

41. Huang, X., Lai, Y., Hang, Z. H., Zheng, H. & Chan, C. T. Dirac cones induced by accidental degeneracy in photonic crystals and zero-refractive-index materials. *Nat. Mater.* **10,** 582–586 (2011).

42. Silveirinha, M. & Engheta, N. Design of matched zero-index metamaterials using nonmagnetic inclusions in epsilon-near-zero media. *Phys. Rev. B* **75,** 075119 (2007).

43. Liberal, I. & Engheta, N. Near-zero refractive index photonics. *Nat. Photonics* **11,** 149–158 (2017).

44. Byrnes, S. J., Khorasaninejad, M. & Capasso, F. High-quality-factor planar optical cavities with laterally stopped, slowed, or reversed light. (2002). doi:10.1364/OE.24.018399

45. Lallier, E. Rare-earth-doped glass and LiNbO_3 waveguide lasers and optical amplifiers. *Appl. Opt.* **31,** 5276 (1992).

46. Silveirinha, M. G. & Engheta, N. Theory of supercoupling, squeezing wave energy, and field confinement in narrow channels and tight bends using ε near-zero metamaterials. *Phys. Rev. B* **76,** 245109 (2007).

47. Joannopoulos, J. D. & Johnson, S. G. Block-iterative frequency-domain methods for Maxwell's equations in a planewave basis. *Opt. Express, Vol. 8, Issue 3, pp. 173-190* **8,** 173–190 (2001).

48. Oskooi, A., Roundy, D., Ibanescu, M., … P. B.-C. P. & 2010, undefined. MEEP: A flexible free-software package for electromagnetic simulations by the FDTD method. *Elsevier*

49. Thijssen, J. Characterization of photonic colloidal crystals in real and reciprocal space. (2007).

50. Joannopoulos, J., Johnson, S., Winn, J. & Meade, R. *Photonic crystals: molding the flow of light.* (2011).

51. Sakurai, J. J. & Commins, E. D. Modern Quantum Mechanics, Revised Edition. *Am. J. Phys.* **63,** 93–95 (1995).